\documentclass[letterpaper, 10 pt]{article}
\usepackage[margin=1in]{geometry}
\usepackage{times}  
\usepackage{cite}
\usepackage{amsmath,amssymb,amsfonts,bm}
\usepackage{algorithmic}
\usepackage{graphicx}
\usepackage{textcomp}
\usepackage{xcolor}
\usepackage{mathtools}
\usepackage{caption,makecell,multirow} 
\usepackage{booktabs} 
\usepackage{hyperref}
\usepackage{authblk} 

\usepackage{tabularx}
\usepackage{adjustbox}
\usepackage{subcaption}
\usepackage{threeparttable}

\usepackage{fancyhdr}

\title{\Large \bf AI-Driven SEEG Channel Ranking for Epileptogenic Zone Localization}
\author[1]{Saeed Hashemi}
\author[1]{Genchang Peng}
\author[1]{Mehrdad Nourani\thanks{Corresponding author: mehrdad.nourani@utdallas.edu}}
\author[2]{Omar Nofal}
\author[2]{Jay Harvey}

\affil[1]{Department of Electrical and Computer Engineering, University of Texas at Dallas, Richardson, TX 75080, USA\\
\texttt{\{SeyedSaeed.HashemiEsmaeilabad, gxp170004, Nourani\}@utdallas.edu}}
\affil[2]{Department of Neurology and Neurotherapeutics, University of Texas Southwestern Medical Center, Dallas, TX 75390, USA\\
\texttt{\{Omar.Nofal, Jay.Harvey\}@UTSouthwestern.edu}}

\begin{document}
\maketitle
\thispagestyle{fancy}
\pagestyle{fancy}
\begin{abstract}
Stereo-electroencephalography (SEEG) is an invasive technique to implant depth electrodes and collect data for pre-surgery evaluation. Visual inspection of signals recorded from hundreds of channels is time consuming and inefficient. We propose a machine learning approach to rank the impactful channels by incorporating clinician's selection and computational finding. A classification model using XGBoost is trained to learn the discriminative features of each channel during ictal periods. Then, the SHapley Additive exPlanations (SHAP) scoring is utilized to rank SEEG channels based on their contribution to seizures. A channel extension strategy is also incorporated to expand the search space and identify suspicious epileptogenic zones beyond those selected by clinicians. For validation, SEEG data for five patients were analyzed showing promising in terms of accuracy, consistency and explainability.
\end{abstract}

\section{Introduction}
\subsection{Motivation}
Epileptic seizures affect millions of people worldwide, with over 30\% are resistant to at least two anti-seizure drugs~\cite{wissel2024early}. For these drug-resistant patients, epilepsy surgery is usually recommended tosuppress seizure occurrences and render the patient seizure-freedom. Typical surgical options include temporal lobe resection, laser interstitial thermal therapy (LITT), and responsive neurostimulation (RNS)~\cite{gavvala2022stereotactic,my2022bibe}. While the principles and approaches are different, the success of epilepsy surgery operations is highly dependent on the accurate localization of the epileptogenic zone (EZ), which is the hypothesized area needed to be surgically eliminated or intervened~\cite{isnard2018french,bartolomei2017defining}. Incomplete or imprecise localization of EZs can lead to unsuccessful surgery outcomes (e.g., seizure recurrence) or even post-surgical implications (e.g., memory deficit), which highlight the critical need for precise targeting of surgical regions during pre-surgical evaluation~\cite{Fisher2014}. \par
Stereo-electroencephalography (SEEG) is a minimally-invasive technique using implantable depth electrodes to collect deep brain activities with high resolution and safety~\cite{abou2018increased}. Each SEEG electrode has 6-16 channels (i.e., contact points) that record neural activities in the implanted region~\cite{mullin2016seeg}. SEEG has been widely adopted in more than 190 medical institutions throughout the USA for EZ localization and pathology evaluation during pre-surgical stages~\cite{youngerman2019}. With multiple depth electrodes implanted on each hemisphere, SEEG involve hundreds of signal channels, leading to insufficient and time-consuming visual inspections for clinicians~\cite{xiao2021automatic}. Also, the manual inspection relies on the clinicians' expertise and experience, making the evaluation subjective and prone to variability. In addition, due to the complex nature of ictal activities, a doctor's manual selection may overlook the underlying regions that contribute to seizure initiation and propagation. To overcome these challenges, machine learning techniques have been applied  to offer data-driven approaches in pre-surgical evaluation, including: (i) EZ localization and evaluation~\cite{paulo2022seeg}, (ii) seizure prediction~\cite{wang2020seizure}, (iii) surgical planning and outcome prediction~\cite{my2023cbm}. Despite these advances, prior works share common limitations: insufficient focus on patient-specific patterns, and minimal integration of clinician feedback. These shortcomings underscore the need for an explainable, data-driven framework that bridges the gap between machine learning advancements and clinical applicability. 

\subsection{Main Contribution}
We propose a machine learning methodology to identify impactful channels for EZ localization during pre-surgical evaluation. Our strategy is based on incorporating clinician's selections and computational finding. We trained a binary classification model using XGBoost\cite{chen2016xgboost} to learn the effective features from ictal periods. We then quantify the impact of SEEG channels (contact points) during seizure events using SHAP (SHapley Additive exPlanations)\cite{lundberg2017unified} based on their contributions to seizures. We propose two analyses: (i) electrode extension and (ii) zone extension, to explore existence of additional EZ regions beyond those identified by clinicians. By incorporating these additional points for data learning, we identify potential EZ regions that might have been overlooked by physicians and discuss the consistency and accuracy for a few real-world epilepsy patients.

This paper is structured as follows. Section~\ref{sec:metho} describes the methodology, including model training and quantification and ranking of contact points. Section~\ref{sec:results} presents the experimental results and their implications for clinical practice. Finally, Section~\ref{sec:conc} concludes with key takeaways and future directions.

\begin{figure*}[tp]
\centering
\includegraphics[width=0.95\textwidth]{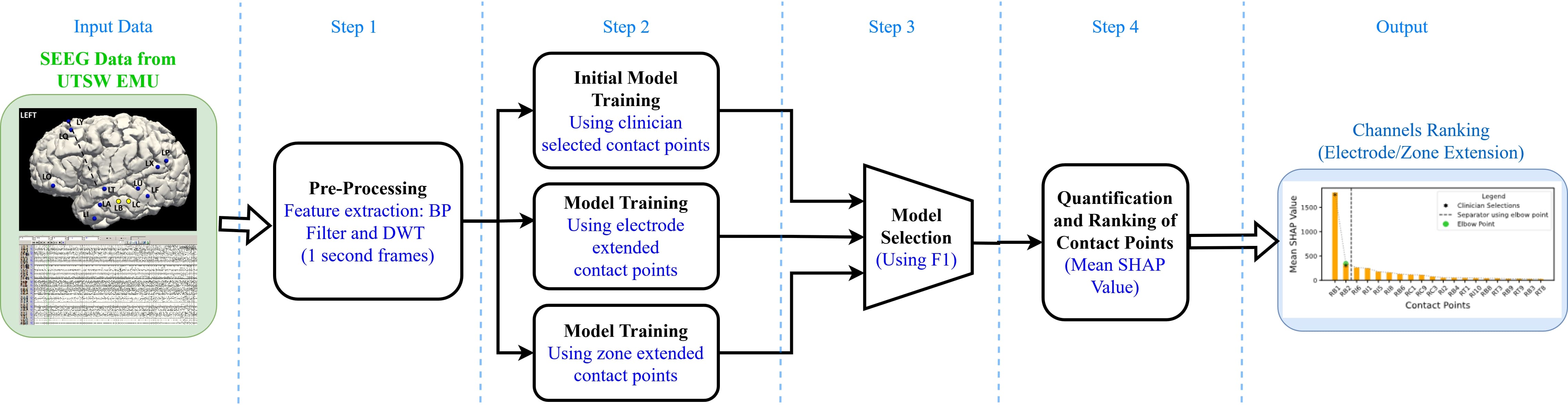}
\caption{Flowchart of proposed methodology.}
\label{fig:flowchart}
\end{figure*}

\begin{figure}[tp]
\centering
\includegraphics[width=0.48\textwidth]{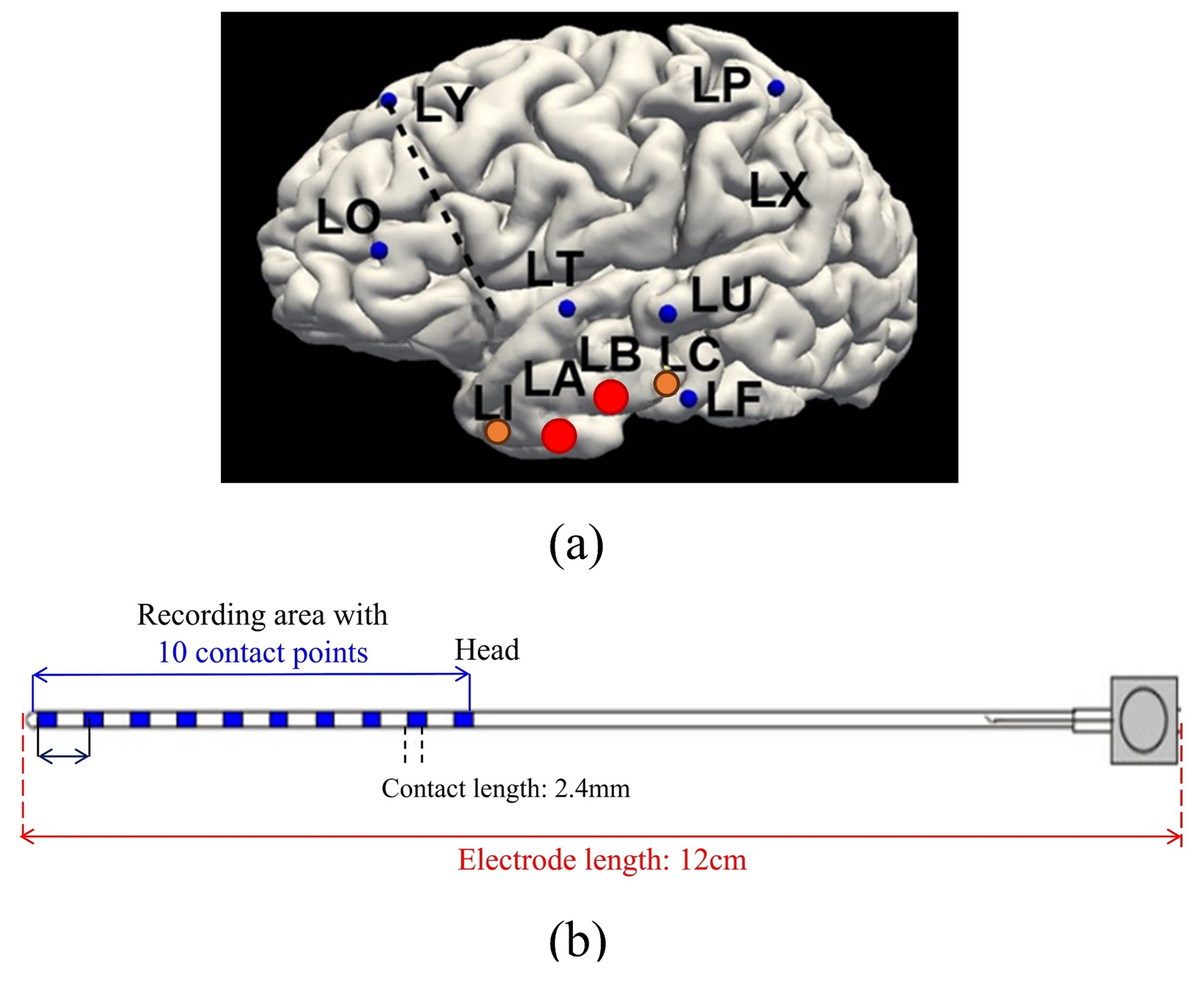}
\caption{SEEG example: (a) Implantation map showing 11 electrodes, with the LA and LB electrodes (clinician-selected) highlighted in red; (b) the 10 contact points located along the LA electrode.}\label{fig:seeg exam}
\end{figure}

\section{Methodology}\label{sec:metho}
The overall methodology is summarized in Figure~\ref{fig:flowchart}, which illustrates the flow of data collection, pre-processing, classification, and ranking method.

\subsection{Step 1: Pre-Processing} \label{pre-processing}
\begin{figure}[tp]
\centering
\includegraphics[width=0.475\textwidth]{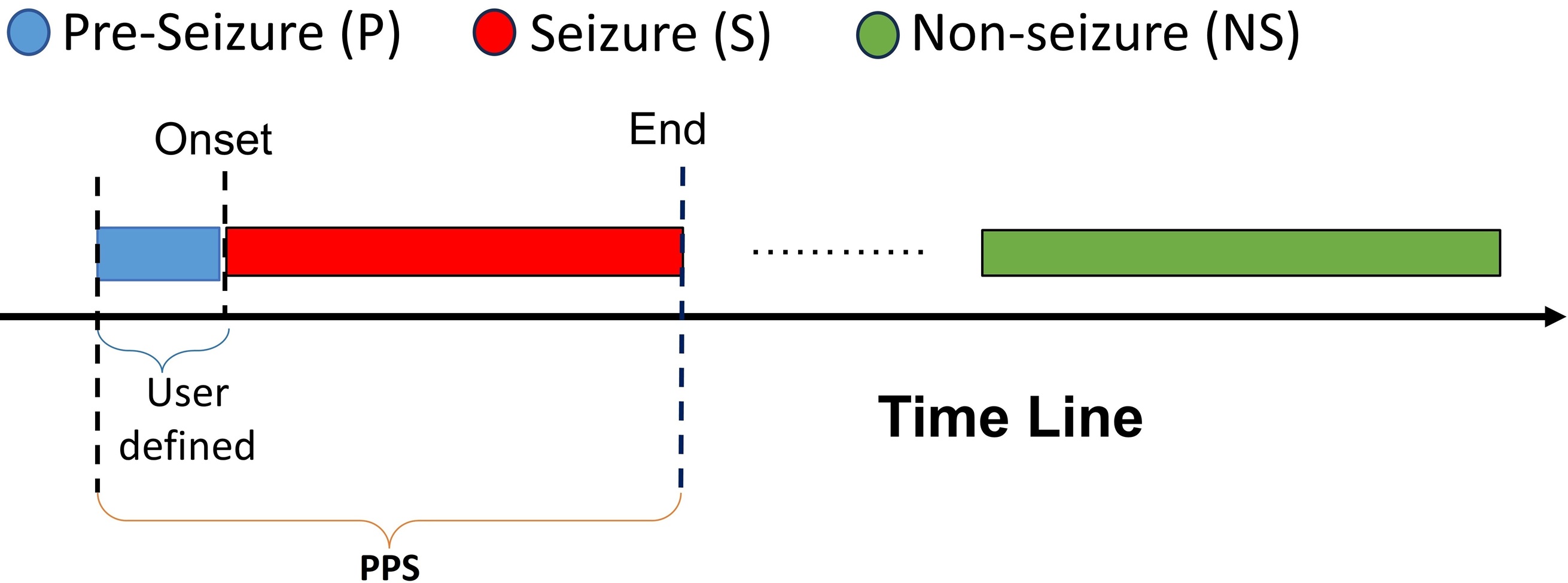}
\caption{Defining two classes for SEEG data.}
\label{fig:Seizure_classes}
\end{figure}

Figure~\ref{fig:seeg exam} shows an example of SEEG that we are going to investigate next, where Figure~\ref{fig:seeg exam}(a) is the electrode implantation map, and Figure~\ref{fig:seeg exam}(b) shows the 10 contact points (i.e., signal channels) from one electrode. In each SEEG channel, 
feature extraction was conducted using a Butterworth bandpass filter (1–60 Hz) to remove noise, followed by Discrete Wavelet Transform (DWT) for frequency-domain analysis. The SEEG data were segmented into 1-second frames with 50\% overlap. To ensure the inclusion of seizure-related information, the seizure period was extended by an additional 20 seconds (a user-defined parameter) from the pre-seizure duration preceding seizure onset. This extended period, referred as the Pre-seizure Plus Seizure (PPS) class, accounts for the challenges of accurately annotating the onset of the seizure and ensures the inclusion of early onset data for comprehensive analysis (Figure~\ref{fig:Seizure_classes}). The binary classification task was conducted using PPS and non-seizure data to analyze the contribution of SEEG contact points systematically.

\begin{figure}[tp]
\centering
\includegraphics[width=0.80\linewidth]{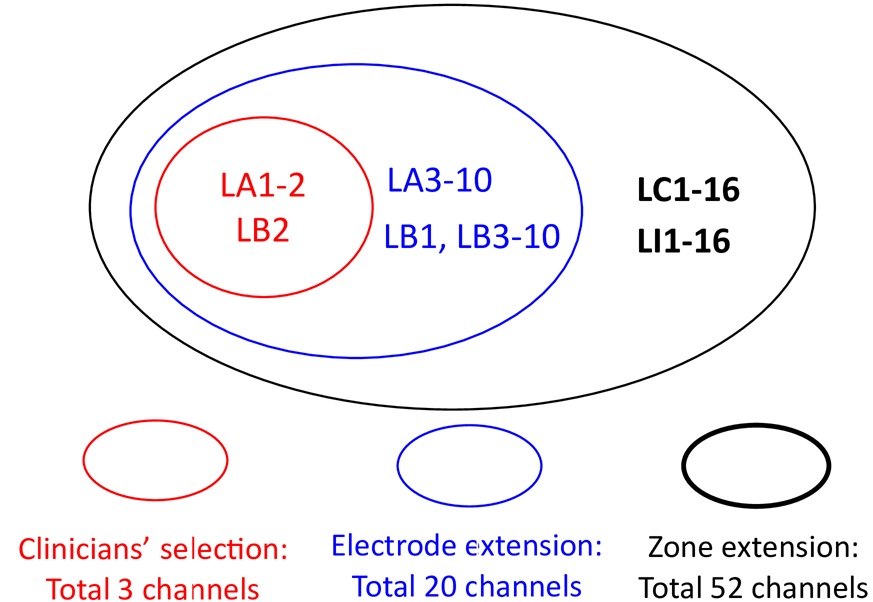}
\caption{Illustration of electrode extension (blue) and zone extension (black). }
\label{fig:extension}

\end{figure}

\subsection{Step 2: Model Training}
An XGBoost classifier \cite{chen2016xgboost}, a scalable and efficient gradient boosting method based on decision trees, was trained to distinguish between PPS and non-seizure data, leveraging its robustness and ability to handle high-dimensional SEEG data. The F1-score serves as a benchmark to assess how effectively the model learns seizure dynamics from clinician-selected contact points. A higher F1-score indicates that these contact points sufficiently capture seizure-related activity, allowing us to trust the model’s ability to analyze their effects. Conversely, a low F1-score suggests that the selected points may lack crucial information or may not be sufficiently informative to characterize seizure activity.

In either case, to enhance seizure localization with broader selection, we apply extension methods to incorporate additional contact points and reassess the F1-score to evaluate potential improvements beyond clinicians selection (denoted in red from Figure~\ref{fig:extension}):
\begin{itemize}
\item \textbf{Electrode Extension}: Incorporates all other contact points within the same electrode (e.g. 10 channels in LA and 10 in LB) beyond clinician-selected points (e.g., LA1-2. LB2), ensuring that potentially relevant seizure activities within the same physical region is considered. In this case, after electrode extension, there are total 20 channels (10 LA and 10 LB) to be analyzed.
\item \textbf{Zone Extension}: Expands the analysis to other adjacent electrodes (e.g. LI and LC in Figure~\ref{fig:seeg exam}(a)) to account for potentially overlooked regions where seizure activity may propagate, allowing a broader assessment of seizure dynamics. In this case, after Zone extension, there are total 52 channels (16 LC, 16 LB, plus 20 channels from electrode extension) to be analyzed. The actual number of analyzed contact points may vary for each patient depending on individual electrode implantation.
\end{itemize}

After applying these extensions, we compare the F1-score to determine whether the newly included contact points provide additional insights into seizure-related activity, as reflected by improved classification performance. This iterative approach ensures a comprehensive and refined understanding of seizure propagation while optimizing the balance between model accuracy and minimizing unnecessary contact points.

\subsection{Step 3: Quantification and Ranking of Contact Points}

To quantify the impact of contact points in our AI-driven ranking method, we apply SHAP, which is an interpretable metric based on machine learning model outputs~\cite{lundberg2017unified}. SHAP assigns each feature (i.e., contact point) a contribution value that reflects its impact on the model’s learning outcomes. By considering all possible subsets of features, SHAP ensures a consistent and comprehensive evaluation of feature importance, providing both local and global interpretability~\cite{latifi2023}.
We define the set of contact points as \( N \), with \( c \) representing the $c$-th contact point in \( N \). The SHAP value \( \phi_{c} \) quantifies the contribution of contact point \( c \):

\begin{equation}
\phi_{c} = \sum_{S \subseteq N \setminus \{c\}} W_{S,c} \Delta_{S,c},
\label{eq:shap_value}
\end{equation}
Where $S$ is a feature (contact point)~\textit{subset} that excludes $c$-th contact point. Then $\Delta_{S,c}$, the marginal contribution of \( c \), is defined as the difference in the model’s output when \( c \) is added into subset \( S \):

\begin{equation}
\Delta_{S,c} = f(S \cup \{c\}) - f(S),
\label{eq:marginal_contribution}
\end{equation}

Here, \( f \) represents the raw output of our classification model before the final layer, which produces values ranging from \(-\infty\) to \(+\infty\) . A negative value indicates that the feature contributes to classifying the input as a non-seizure instance, whereas a positive value signifies its contribution toward classifying the input as PPS class. The weighting factor \( W_{S,c} \), the weighting factor in Equation~\ref{eq:shap_value}, adjusts the influence of subsets based on the subset size $|S|$:
\begin{equation}
W_{S,c} = \frac{|S|!(|N| - |S| - 1)!}{|N|!}.
\label{eq:weighting_factor}
\end{equation}

While the SHAP computation involves evaluating all possible subsets, leading to an exponential growth of \( 2^{|N| - 1} \) subsets, the computational cost can be efficiently managed in specific scenarios. For instance, when using tree-based models like XGBoost, the SHAP algorithm optimizes these calculations by leveraging the structure of decision trees, significantly reducing the computational burden\cite{lundberg2018consistent}. Furthermore, as this study analyzes SEEG data locally, focusing only on PPS
data from limited contact points (e.g., $N<20$), the process becomes computationally efficient and feasible even with limited resources. \par

We use patient 1000 as an example to illustrate the procedure of SHAP calculation, where the clinician-selected contact points were LA1-3, LB1-2, and LC1-2, leading to $N=8$ points. An XGBoost binary classifier \( f \), trained on PPS and non-seizure data, was used to calculate SHAP values for contact point LA1 on PPS data. To calculate the SHAP score $\phi_{\rm{LA1}}$ of LA1, the model output \( f \) was evaluated with and without LA1 across all subsets \( S \subseteq N \setminus \{\text{LA1}\} \), and the marginal contributions were computed as described in Equation~\ref{eq:marginal_contribution}. \par
In our experiment, to summarize the SHAP score of one contact point over the entire ictal events (i.e., PPS periods), we calculate the SHAP score for every $t$-th frame,(\( t = 0 \) to \( t = T \)), resulting in a sequence of SHAP values for LA1:

\begin{equation}
\phi_{\rm{LA1}}^{(0)}, \phi_{\rm{LA1}}^{(1)}, \ldots, \phi_{\rm{LA1}}^{(t)}, \phi_{\rm{LA1}}^{(T)}.
\label{eq:shap_values_la1}
\end{equation}\par
Then the \textit{mean} SHAP value of LA1 $\Phi_{\rm LA1}$ is calculated over all $T$ frames in Equation~\ref{eq:shap_values_la1} to quantify its holistic importance of across the entire PPS data:

\begin{equation}
\Phi_{\rm LA1} = \frac{1}{T} \sum_{t=1}^{T} \phi_{\rm{LA1}}^{(t)}.
\label{eq:mean_shap_value}
\end{equation}

In the final step, we applied the \textit{elbow method} to identify the most impactful contact points in distinguishing seizure. By computing the second derivative of the mean SHAP values, we identified the \textit{elbow point} \( k^* \), where the decline in importance slows significantly:

\begin{equation}
k^* = \arg \max_k \left( \frac{\partial^2 \Phi_c}{\partial k^2} \right),
\label{eq:elbow_point}
\end{equation}

Here, \( k \) represents the index of contact points ranked by their mean SHAP values \( \Phi_c \), and \( k^* \) denotes the elbow point, indicating the index at which the rate of change in \( \Phi_c \) is maximized. We then select the top contact points up to \( k^* \), ensuring that only the most influential ones were retained for analysis.

\par

\section{Experimental Results}\label{sec:results}
This section presents the evaluation of classification performance and effectiveness of extended SEEG contact point analysis in identifying epileptogenic zones for surgical interventions, such as Responsive Neurostimulation (RNS) or Laser Interstitial Thermal Therapy (LITT). 
\subsection{SEEG Dataset}
This study is conducted under Institutional Review Board (IRB) IRB-21-198 approved by the University of Texas at Dallas (UTD) and the University of Texas Southwestern Medical Center (UTSW). We analyze SEEG data from five unilateral epilepsy patients who underwent LITT at UTSW (Table~\ref{table:patient_info}). SEEG data was recorded using the Nihon Kohden EEG 1200 system at a sampling rate of 1000 Hz, ensuring high-resolution signals. Each patient’s SEEG data includes up to 16 contact points per electrode and over 150 total contact points, varying by case. Since the input data and configuration differ for each patient, the analysis is personalized to ensure both accurate classification performance and clinically-meaningful contact point selections. The objective is to quantify SEEG contact points based on their contribution during seizures, helping clinicians identify the most informative points and prioritize overlooked data. This ranked quantification supports better clinical decisions and analysis. \par

\begin{table}[ht]
\centering
\caption{Patient information and clinician-identified seizure onset points.}
\label{table:patient_info}
\scalebox{0.9}{
\begin{tabular}{@{}lccccc@{}}
\toprule
Patient & \multirow{2}{*}{Age} & \multirow{2}{*}{Gender} & Total Seizure & Contact &  Onset \\ 
ID &   &   &   Duration (sec.)  &  Points No. (\#)  & Points  \\
\midrule
\multirow{2}{*}{1000} & \multirow{2}{*}{37} & \multirow{2}{*}{Male}   & \multirow{2}{*}{342.77 s} & \multirow{2}{*}{175} & LA1-3, LB1-2, \\
  &  &  &  & & LC1-2 \\
  \cmidrule{1-6}
1300 & 50 & Female & 147 s    & 183 &  \
LB1-2, LC 1-2,\\
&  &  &  & & LY 1-6\\ 
 \cmidrule{1-6}
1500 & 52 & Female & 200 s    & 217 & LB1-2 \\ 
 \cmidrule{1-6}
1600 & 43 & Female & 453 s    & 213 & RB1-2 \\ 
 \cmidrule{1-6}
1700 & 21 & Male   & 204 s    & 207 & LB1-2 \\ 
\bottomrule
\end{tabular}}
\end{table}
\subsection{Experimental Setup}

An XGBoost classifier was trained to learn seizure dynamics, distinguishing Pre-Seizure Plus Seizure (PPS) from Non-Seizure states. While XGBoost was chosen for its speed and interpretability, other decision-tree-based or statistical models can also be utilized. To ensure robust and reliable results, we employed 5-fold cross-validation and split the data into 80\% for training and 20\% for testing.

\subsection{Model Performance and Impact of Extensions}

Our model performance was assessed using F1-scores (i.e. $ \frac{2 \times \text{Precision} \times \text{Recall}}{\text{Precision} + \text{Recall}}$), starting with clinician-selected contacts as a reference and then expanding the contact points using electrode and zone extensions (Table~\ref{tab:performance}). \par
\begin{table}[ht]
    \centering
    \caption{Effect of electrode and zone extensions on model performance}
    \label{tab:performance}
    \scalebox{0.95}{
    \begin{tabular}{@{}cccc@{}}
 
        \toprule
        \multirow{2}{*}{Patient ID} & \multicolumn{3}{c}{Average F1-score (\%)} \\ 
        \cmidrule(lr){2-4}
         & Clinician Selected & Electrode Extended & Zone Extended\\
        \midrule
                1000 & 93\% & 95\% & 96\% \\
                1300 & 95\% & 95\% & 94\% \\
                1500 & 81\% & 89\% & 92\% \\
                1600 & 87\% & 94\% & 95\% \\
                1700 & 95\% & 98\% & 98\% \\
                \bottomrule
                
    \end{tabular}}
\end{table}
In most cases, extensions improved performance, As observed in patient 1600, where electrode extensions increased the F1-score, suggesting additional seizure-related patterns in neighboring contact points. Zone extensions further enhanced results in cases like patient 1500, where a notable performance boost indicated seizure activity beyond the initially selected contacts. However, in some patients, like patient 1300, scores remained stable, suggesting that clinician-selected contacts were already optimum choice. These results highlight the variability in seizure patterns across patients and demonstrate the model’s ability to refine seizure dynamics analysis through targeted extensions.

These results confirm that the model successfully identifies seizure dynamics and that expanding contact point selection can uncover additional meaningful features, providing a broader and more precise framework for seizure analysis.

\begin{figure}[t]
\centering
\includegraphics[width=0.485\textwidth, height=0.485\textwidth]{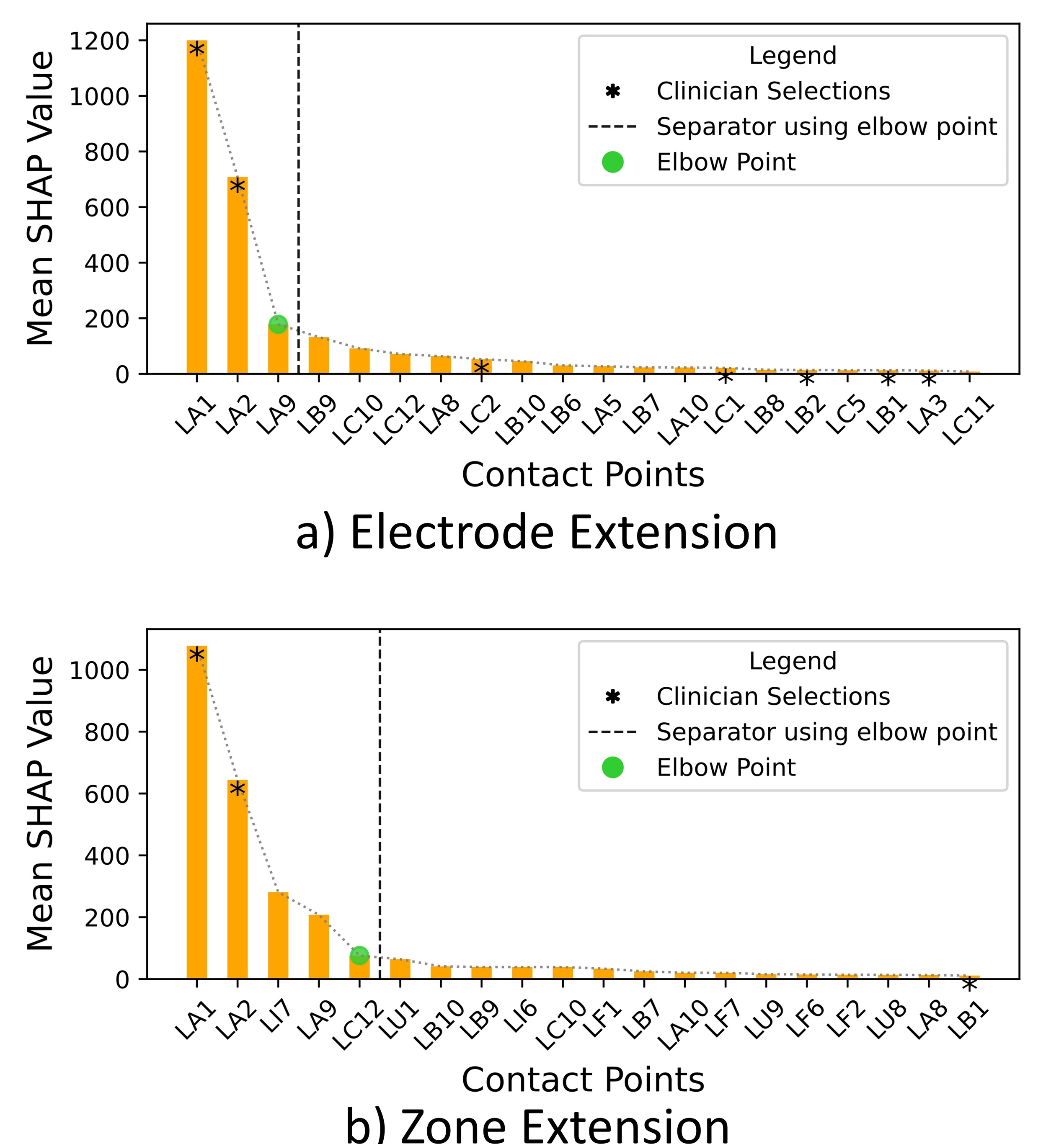}
\caption{SHAP-based ranking for patient 1000. (a) Electrode extension, where LA9 is identified as a new finding. (b) Zone extension, where LI7, LA9, and LC12 are newly detected contact points.}
\label{fig:p1000}
\end{figure}

\begin{figure}[tp]
\centering
\includegraphics[width=0.485\textwidth,height=0.485\textwidth]{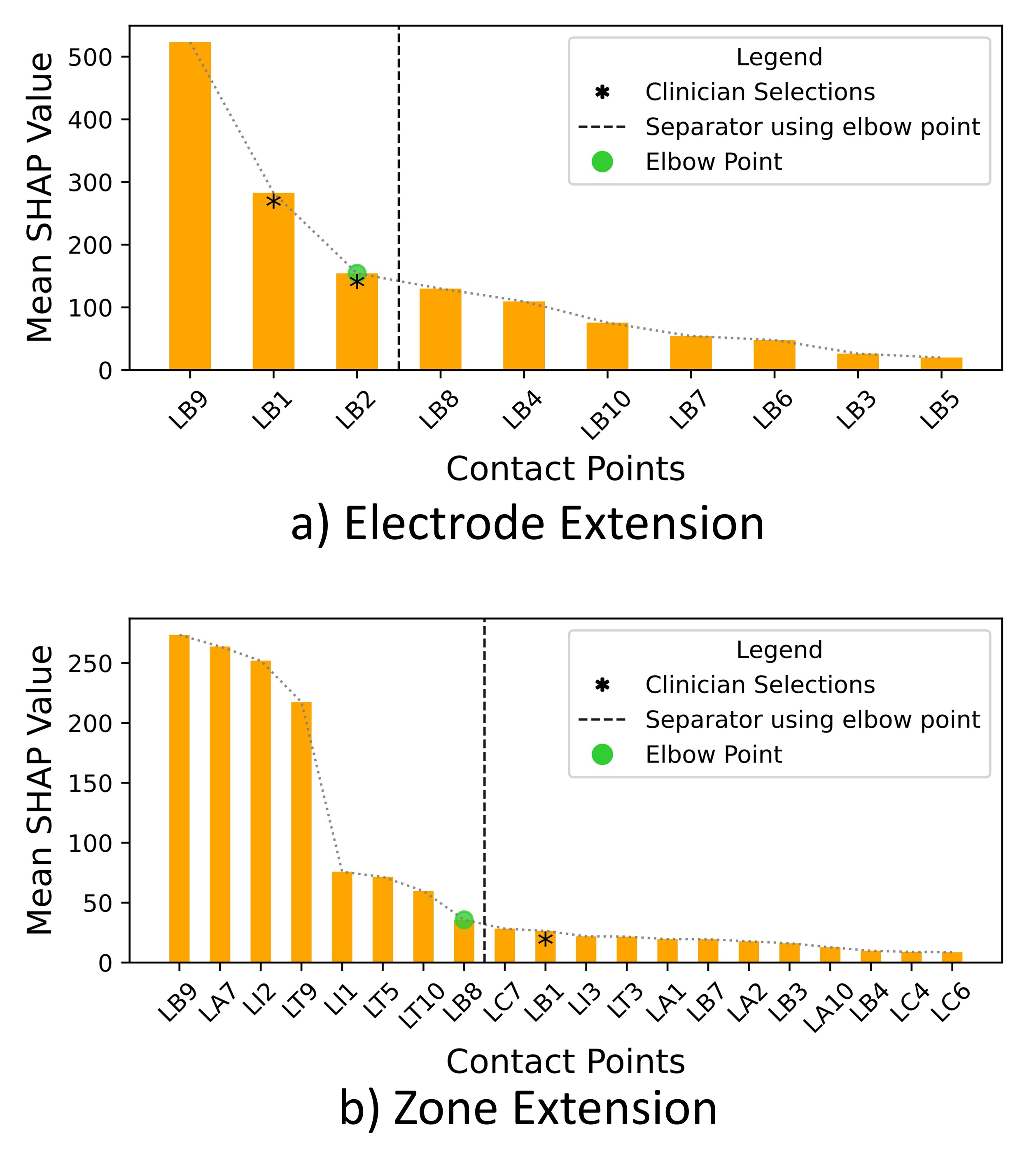}
\caption{SHAP-based ranking for patient 1500. (a) Electrode extension, where LB9 is identified as a new finding. (b) Zone extension, where LB9, LA7, LI2, LT9, LI1, LT5, LT10, and LB8 are newly detected contact points.}
\label{fig:p1500}
\end{figure}

\subsection{Data-Driven Insights}

\begin{figure*}[tp]
\centering
\includegraphics[width=.98\textwidth]{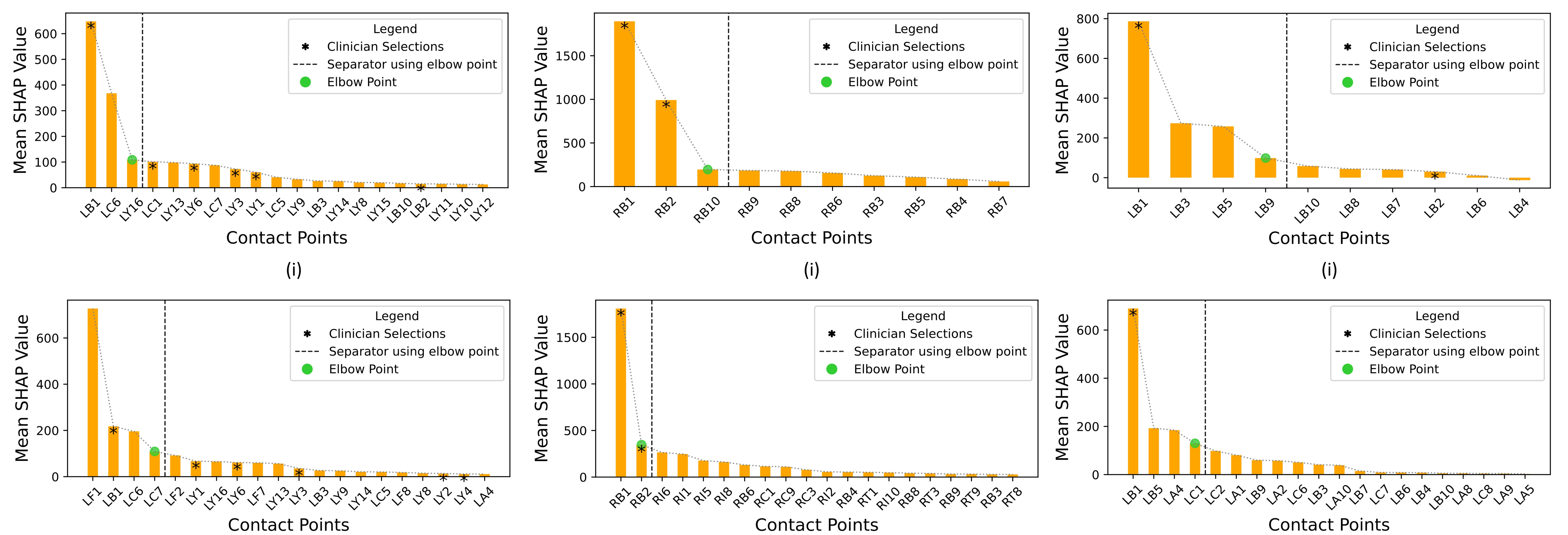}
\caption{SHAP-based ranking for patient 1300, 1600, and 1700. Sub-figures labeled (i) refer to electrode extension rankings, while sub-figures labeled (ii) show zone extension rankings. For better visualization, only the top 20 ranked contact points are displayed in the bar plots.}
\label{fig:rest_patients}
\end{figure*}

Our data-driven method demonstrates that electrode and zone extensions identify high-impact SEEG contact points, complementing clinician expertise (Table \ref{tab:shap_rankings}). In most cases, a subset of clinician-selected contact points remains among the top-ranked points, confirming the model’s ability to highlight critical clinician-selected points. Furthermore, the extensions reveal new high-impact contact points that may have been previously overlooked. For example, in patient 1000 (Figure \ref{fig:p1000}), the zone extension identifies LI7, LA9, and LC12—contact points exhibiting additional contributions and displaying seizure-related patterns (Figure \ref{fig:p1000}(b)).

In another case, patient 1500, our method first identified LB9 during the electrode extension phase as having a greater impact than the clinician-selected contacts (Figure \ref{fig:p1500}(a)). Subsequently, the zone extension for this patient revealed a distinct set of high-impact contact points, none of which had been originally selected by the clinician, emphasizing the value of systematic AI computational models in complex cases (Figure \ref{fig:p1500}(b)). 
These findings highlight two key insights. First, the identification of more impactful contact points by data analytics that clinicians can consider. Second, evidence that seizure activity propagates to neighboring electrodes, revealing additional critical contact points that further illustrate seizure complexity in this case. These findings assist clinicians to evaluate adjacent contact points to gain a more comprehensive understanding of seizure dynamics.

Beyond these cases, the SHAP ranking results for the remaining patients further confirm these trends, as shown in Figure \ref{fig:rest_patients}, where high-impact contact points consistently emerge across different cases. This underscores the model’s ability to dig out useful knowledge across patients, effectively identifying critical contact points that might have been hard (or impossible) to find during traditional visual inspection.

While the classification results (i.e. F1-scores) for retrospective SEEG data indicate the advantage of our AI-driven analysis, a more comprehensive validation is needed. In particular, the informativeness and accuracy of the high-impact contact points, found by electrode/zone extensions, need to be validated. As future work, we plan to validate this aspect of the methodology by sharing the AI finding with clinicians {\em during} the pre-surgery evaluation and surveying their views as decisions being made.

\begin{table}[ht]
    \centering
    
    \caption{Comparison of clinician selections and model-identified high-impact contacts}
    \label{tab:shap_rankings}
    \scalebox{0.93}{
    \begin{threeparttable}
    \begin{tabular}{@{}cccc}
        
        \toprule
        Patient ID & Clinician Selected & Electrode Extension & Zone Extension \\
        \midrule
        \multirow{2}{*}{1000}  & LA1-3, LB1-2, & LA1*, LA2*,  & LA1*, LA2*, LI7,  \\
            &                    LC1-2  &  LA9, LB9  & LA9, LC12 \\
\midrule
        \multirow{2}{*}{1300}  & LA1-4, LI1-4,  & LB1*, LC6,  & LF1, LB1*,  \\
         &                      LC1, LZ1-3&LY16, LC1&LC6\\
         \midrule
        \multirow{3}{*}{1500}  & LB1-2 & LB9, LB1*, & LB9, LA7, LI2,  \\
        &                           & LB2*&LT9, LI1, LT5,\\
        &&&LT10, LB8\\
        \midrule
        \multirow{2}{*}{1600}  & RB1-2 & RB1*, RB2*, & RB1*, RB2*, RI6 \\
        &   &   RB10    &   \\
        \midrule
        \multirow{2}{*}{1700}  & LB1-2 & LB1*, LB3, & LB1*, LB5, LA4,  \\
        &&LB5, LB9&LC1\\
        \bottomrule
    \end{tabular}
    \begin{tablenotes}
            \item * clinician-selected contact points
        \end{tablenotes}
    \end{threeparttable}
    \vspace{0.2cm}
    
    }
\end{table}

\section{Conclusion} \label{sec:conc}

The proposed AI-driven framework effectively captures seizure dynamics by identifying high-impact SEEG contact points, often aligning with clinician-selected regions while also revealing additional seizure-related points. Electrode and zone extensions enhanced model performance in most cases, underscoring the importance of broader spatial evaluation in seizure characterization. However, patient-specific variability highlights the complexity of epilepsy and the need for adaptive analytical strategies. While this study was conducted on a limited dataset, future research should assess the method’s generalizability on a larger and more diverse patient cohort, including cases with bilateral seizure onsets. With additional utilizing deep learning techniques could further refine feature extraction and classification.

\end{document}